\documentclass[twocolumn,aps,prb,amssymb,showpacs,groupedaddress,superscriptaddress]{revtex4-1}

\usepackage{amsfonts}
\usepackage{txfonts}
\usepackage{graphicx}
\usepackage{subfigure}%
\usepackage{bm}%

\begin{document}

\title{Floquet Topological Superfluid and Majorana Zero Modes in Two-Dimensional Periodically Driven Fermi Systems}

\author{Xiaosen Yang}
\email{yangxs@csrc.ac.cn}
\affiliation{Beijing Computational Science Research Center,Beijing, 100084, China}
\affiliation{Department of physics, Jiangsu University, Zhenjiang, 212013, China}

\date{\today}

\begin{abstract}
We propose a simple approach to realize two-dimensional Floquet topological superfluid by periodically tuning the depth of square optical lattice potentials. We show that the periodic driving can induce topological phase transitions between trivial superfluid and Floquet topological superfluid. For this systems we verify the anomalous bulk-boundary correspondence, namely that the robust chiral Floquet edge states can appear even when the winding number of all the bulk Floquet bands is zero. We establish the existence of two Floquet Majorana zero modes separated in the quasienergy space,  with $\varepsilon_{0,\pi}=0,\pi/T$ at the topological defects.
\end{abstract}

\date{\today}
\pacs{03.67.Vf, 71.10.Fd, 74.40.Gh}
\maketitle
\section{Introduction}

In recent years, topological states of matter have attracted much interests in condensed matter and cold atom physics\cite{HasanRevModPhys2010,XLQiRevModPhys2011}. These quantum states are distinguished by topological invariants\cite{thouless1982,kane2005b,fu2007b,qi2008,wang2010order,wang2010a,wang2012simplified} instead of order parameters of the Landau theory.
More Recently, investigations of topological matter have been extended to periodically driven quantum systems or Floquet systems\cite{eckardtPhysRevLett2005,TKitagawaPRB2010,NLindnerNATPHY2011, KArijitPRL2013, DEliuPRL2013, LJiangPRL2011, YKatanPRL2013,YHWangScience2013, LGomezPRL2013,rechtsmannature2013,MLababidiPRL2014, satoarxiv2014}.
Compared to static systems, these periodically driven systems enjoy many new and fascinating properties. For instance, Floquet topological insulator can have robust topological edge states even though the Chern numbers of all the quasienergy bands vanish\cite{MSRudnerPRX2013}.  In addition, for Floquet systems there are two particle-hole conjugated energies, namely\cite{LJiangPRL2011} $\varepsilon_{0,\pi}=0,\pi/T$.  Motivated by the rich phenomena in Floquet systems, many novel phases have been proposed in such system, for instance, Floquet topological insulators\cite{NLindnerNATPHY2011,lindnerPhysRevB2013, MSRudnerPRX2013, MRechtsmanNature2013, YKatanPRL2013}, Floquet topological superfluids\cite{LJiangPRL2011, QJTongPRL2013}, Floquet fractional Chern insulator\cite{grushinPhysRevLett2014} and Floquet Weyl semimetal\cite{wangarxiv2014}.

The rich phenomena in Floquet systems have generated considerable interests in realizing them experimentally. Sofar there are several possible routes to realize the Floquet topological phases, including coupling a modulated electromagnetic field to electron in solid states\cite{gomezarxiv2013,grushinPhysRevLett2014,Titumdarxiv2014}, periodically tuning the chemical potential\cite{LJiangPRL2011}, and shaking optical lattice\cite{limPhysRevLett2008,limPhysRevA2010,libertoPhysRevA2011,kogheePhysRevA2012,haukePhysRevLett2012, zhengarxiv2014} in cold atom.



In this paper, we shall study two-dimensional Floquet Fermi systems with cold atoms. The two dimensionality is special in that non-Abelian statistics can be realized here. With the motivation of potential applications in non-Abelian statistics and topological quantum computation, we propose a simple approach to realize two-dimensional Floquet topological superfluids by periodically tuning the depth of optical lattice potentials.

The rest of this paper is organized as follows. We first demonstrate that Floquet topological superfluid phase arises when the driving frequency is lowered below the bandwidth, when robust chiral edge states span the gap at $\varepsilon_{\pi}$. When the frequency if further lowered to below half bandwidth, robust chiral edge states span both the gaps around $\varepsilon_{0,\pi}$, with the Chern number of all the Floquet bands vanishing.
We then show that the Floquet topological nontrivial phases are weak pairing phases and have rather clear Fermi surfaces, thus the topological phase transition is a transition from the strong pairing phase to the weak pairing phase. Finally, we show that two flavor Floquet Majorana zero modes are localized at the topological defects in Floquet topological superfluid, moreover, the two flavors Floquet Majorana zero modes are separated at quasienergy space.

\section{Periodic Driving on the hopping}

We consider a spin-polarized Fermi gas loaded on square optical lattice potentials $V(x,y)=V(\sin ^2 k_{0}x +\sin^2 k_{0}y)$. For deep optical lattice potentials and low temperature case, Fermions are restricted to the lowest vibrational level at each site. The kinetic energy of the atoms are frozen except for the tunneling between the nearest neighboring sites. The nearest neighbor tunneling amplitude $J_0$ is determined by $J_{0}/E_r \simeq \frac{4}{\sqrt{\pi}}\left( \frac{V}{E_{r}}\right)^{3/4}\exp \left[-2\left(\frac{V}{E_{r}}\right)^{1/2}\right]$\cite{peilPhysRevA2003,blochRevModPhys2008,giorginiRevModPhys2008} with recoil energy $E_{r}=\hbar^{2} k_{0}^{2}/2m$ and the wave vector of the laser light $k_{0}=2\pi/\lambda$ ($\lambda$ is the wavelength of the laser lights). At the mean-field level we can introduce a pairing potential $\Delta({\bf k})$,  and write the superfluid Hamiltonian as
\begin{eqnarray}
H=\sum_{{\bf k}}\varepsilon_{{\bf k}}c^{\dagger}_{{\bf k}}c_{{\bf k}} + \sum_{{\bf k}} [\Delta({\bf k})c^{\dagger}_{{\bf k}}c^{\dagger}_{-{\bf k}} + H.c], \label{Hamiltonian}
\end{eqnarray}
where $c^{\dagger}_{{\bf k}}(c_{{\bf k}})$ denotes the creation (annihilation)
operators for fermions with momentum ${\bf k}$, and $\varepsilon_{{\bf k}}=2 J_{0} (2 - \cos k_{x}-\cos k_{y})-\mu$, $\mu$ being the chemical potential. For a spin-polarized (or spinless) Fermi system, the pairing potential is odd under inversion transformation, namely, $\Delta(-{\bf k})=-\Delta({\bf k})$. As a result, the pairing potential has gapless nodes at time-reversal invariant momenta ${\bf k}_{c}=[(0,0),(0,\pi),(\pi,0),(\pi,\pi)]$. For simplicity, we consider a $p$-wave pairing  $\Delta({\bf k})=\Delta(\sin k_{x} - i\sin k_{y})$\cite{NReadPRB2000,BLiuPRA2012,YJHanPRL2009,TLHoPRL2005,iskinPhysRevB2005,massignanPhysRevA2010}.

In the Nambu basis $\psi_{{\bf k}}^{\dag}=(c^{\dagger}_{{\bf k}},c_{{\bf k}})$, the Hamiltonian can be written as $\hat{H}=\frac{1}{2}\sum_{{\bf k}} \psi_{{\bf k}}^{\dag} H({\bf k}) \psi_{{\bf k}}$ with $H({\bf k})=\vec{n}_{{\bf k}} \cdot \boldsymbol{\sigma}$, where $\vec{n}_{{\bf k}}=(\Delta \sin k_x , \Delta \sin k_y , \varepsilon_{{\bf k}})$. The topological properties can be characterized by a winding number, which is given by the following well-known formula
\begin{eqnarray}
W= \frac{1}{8\pi} \int d^{2}k \epsilon_{ij} \hat{n}_{{\bf k}} \cdot (\partial_{k_{i}} \hat{n}_{{\bf k}} \times \partial_{k_{j}} \hat{n}_{{\bf k}}),
\end{eqnarray}
with $\hat{n}_{{\bf k}}=\vec{n}_{{\bf k}}/E_{{\bf k}}$ and $E_{{\bf k}}=|\vec{n}_{{\bf k}}|=\sqrt{\varepsilon_{{\bf k}}^{2}+|\Delta({\bf k})|^{2}}$. The superfluid is topological nontrivial ($W\neq 0$) for $0<\mu<8J_{0}$ and trivial ($W=0$) for $\mu>8J_{0}$ and $\mu<0$.

Now let us tune the depth of the lattice potential with period $T$, as a result, the nearest neighbor tunneling amplitude is periodically varying. For simplicity, we assume that the tunneling amplitude varies like $J(t)=J_{0}-J_{D}\cos (\omega t)$.  Consequentially, the above Hamiltonian becomes periodically time-dependent and satisfies $H(t+T)=H(t)$ with period $T=2\pi/\omega$. The periodic time-dependent Hamiltonian can be rewritten as
\begin{eqnarray}
\hat{H}(t)=\frac{1}{2}\sum_{{\bf k}}\psi_{k}^{\dag} H({\bf k},t) \psi_{k}, \label{totalH}
\end{eqnarray}
where
\begin{eqnarray}
H({\bf k},t)&=&H({\bf k})+H_D({\bf k}) \cos(\omega t)\nonumber\\
   &=&\vec{n}_{{\bf k}} \cdot \boldsymbol{\sigma} + \vec{V}({\bf k}) \cdot \boldsymbol{\sigma} \cos(\omega t),\label{eq5}
\end{eqnarray}
in which $\vec{V}({\bf k})=(0,0,\varepsilon_{{\bf k}}^{D})$ and  $\varepsilon_{{\bf k}}^{D}=-2 J_{D} (\cos k_x +\cos k_y )$.

To study the properties of the driven system, we start from the Schr\"{o}dinger equation\cite{shirleyPhysRev1965}:
\begin{eqnarray}
i \partial_{t} \Psi({\bf k}, t) = H({\bf k},t) \Psi({\bf k},t).
\end{eqnarray}
According to the Floquet theorem, the wave function satisfies $\Psi({\bf k},t) = e^{-i\varepsilon({\bf k})t}\Phi({\bf k},t)$ with Floquet states $\Phi({\bf k},t)$ satisfying $\Phi({\bf k},t)=\Phi({\bf k}, T + t)$ and the Floquet equation $[H({\bf k},t) - i\partial_{t}]\Phi({\bf k},t) = \varepsilon({\bf k}) \Phi({\bf k},t)$. Here the quasienergies $\varepsilon({\bf k})$ are defined modulo $2\pi/T$, which is analogous to the lattice momentum in the Bloch band theory. Therefore, there exist two particle-hole conjugated quasienergies $\varepsilon_{0,\pi}=0, \pi/T$ in the quasienergy spectrum, in contrast to the static systems with only one particle-hole conjugated energy $\varepsilon_{0}$. This feature implies nontrivial effects on the Floquet systems.

The Floquet states can be expanded as $\Phi({\bf k},t)=\sum_{m} \phi_{m}({\bf k}) e^{i m \omega t}$, the coefficient $\phi_{m}$ satisfying\cite{MSRudnerPRX2013}
\begin{eqnarray}
\sum_{m'} H_{m,m'}({\bf k}) \phi_{m'}({\bf k})=\varepsilon({\bf k})\phi_{m}({\bf k}).\label{floqeq}
\end{eqnarray}
in which the time-independent Floquet Hamiltonian $H_{m,m'}({\bf k})$ is given by
\begin{eqnarray}
H_{m,m'}({\bf k})=m \omega \delta_{m,m'} +\frac{1}{T} \int_{0}^{T}dt H({\bf k},t) e^{i(m'-m)\omega t},\label{floqh2}
\end{eqnarray}
with $H_{m,m}({\bf k})=m \omega +H({\bf k})$, $H_{m+1,m}({\bf k})= \frac{1}{2} H_D({\bf k})$ and $H_{m,m+1}({\bf k})= \frac{1}{2} H_D({\bf k}) ^{\dag}$.
The topological properties of the periodically driven superfluid are defined in terms of the time-independent Floquet Hamiltonian in Eq.(\ref{floqh2}).

\section{Floquet Topological Superfluid}

In accordance with the general principle of bulk-boundary correspondence, the topological properties of a bulk system can be characterized by the robust edge states. To investigate the topological properties of the periodically driven systems, we consider the Floquet Hamiltonian in Eq.(\ref{floqh2}) in a strip geometry, in which the edges are along the $x$ direction at $L=0,60$.

\begin{figure}
\includegraphics[width=8.5cm, height=6.5cm]{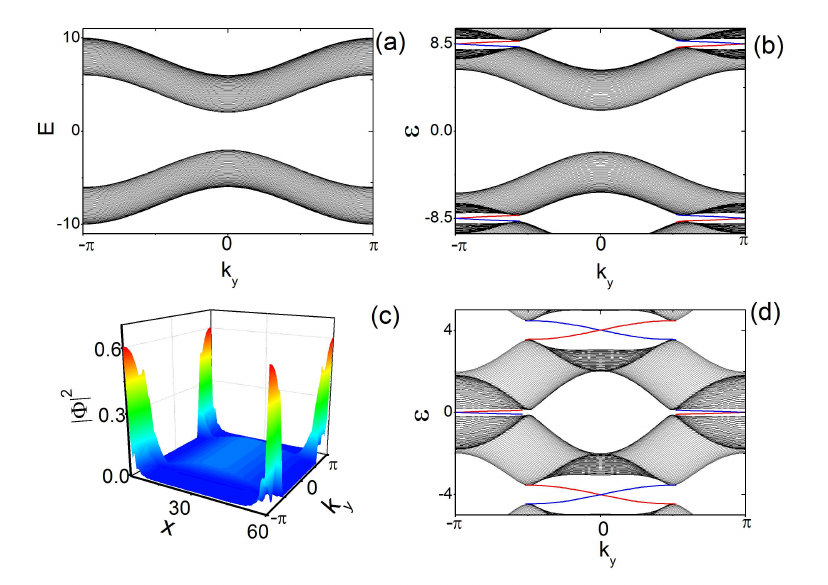}
\caption{The edge states of Flouqet topological superfluid. (a): The spectrums of undriven systems $H(k)$ given by Eq.(\ref{eq5}) for $\mu=-2$, $J_{0}=1$ and $\Delta=1$ in a strip geometry. (b): The spectra of the Floquet Hamiltonian given by Eq.(\ref{floqh2}) in a strip geometry, with $\mu=-2$, $J_{0}=1$, $J_{D}=1$, $\Delta=1$, and  $\omega=17$. (c): The wave function $|\Phi|^{2}$ of the edge states in (b). The pair chiral edge states are localized at two boundaries respectively. (d): The spectra of the Floquet Hamiltonian in a strip geometry, with $\mu=-2$, $J_{0}=1$, $J_{D}=1$, $\Delta=1$, and  $\omega=8$.} \label{ES1}
\end{figure}

As a comparison, first we present numerical results without periodic driving, as shown in Fig.\ref{ES1}(a).  As discussed above, without the driving the superfluid is topological trivial when the chemical potential $\mu>8J_{0}$ or $\mu<0$. Accordingly, there should be no robust edge states at the open boundaries, which is verified in our numerical calculations. On the other hand, when the system is driven with a frequency much greater than other energy scales, there exist a large gap at $\varepsilon_{\pi}$, and the system is again topologically trivial.

Suppose that we gradually decrease the driving frequency.
The gap at $\varepsilon_{\pi}$ can be closed and reopened at $\omega=2W={\rm Max}( 2E_{{\bf k}_{c}}) $, wherein the nearest Floquet bands (up branch of the $m$-th Floquet band and down branch of the $m+1$-th Floquet band) become inverted. When $\omega<2W$, robust chiral Floquet edge states span the gap near $\varepsilon_{\pi}$, as shown in Fig.\ref{ES1}(b).
The chiral Floquet edge states are different from those appearing in static topological superfluid in that edge states span the gap at $\varepsilon_{\pi}$ instead at $\varepsilon_{0}$. A pair chiral Floquet edge states are localized at the two boundaries of the system, which can been found in Fig.\ref{ES1}(c). Therefore, changing driving frequency induces a topological phase transition from a trivial superfluid phase to Floquet topological superfluid phase.

To see the topological phase transition from a more transparent perspective,  we can derive a time-independent effective Hamiltonian $H_{eff}({\bf k})$ by evolution operator over a period $T$:
\begin{eqnarray}
U({\bf k},T)=\textsl{T} \text{e}^{-i \int_{0}^{T} H({\bf k},t) dt}\simeq \text{e}^{-i H_{eff}({\bf k}) T}.
\end{eqnarray}
To derive the effective Hamiltonian for adjacent Floquet bands (between up branch of $m$ Floquet band and down branch of $m+1$ Floquet band), it is convenient to shift the energy of Floquet band by $\pm\omega/2$. Following the method of Ref.\cite{NLindnerNATPHY2011}, we introduce a unitary transformation $O({\bf k},t)= \exp [- i \hat{n}_{\bf k} \cdot  \boldsymbol{\sigma} \omega t /2]$. The transformed effective Hamiltonian is found as $H_{eff}({\bf k})=\vec{n}'_{{\bf k}} \cdot \boldsymbol{\sigma}$ with
\begin{eqnarray}
\vec{n}'_{{\bf k}}= (1-\frac{\omega}{2 E_{\bf k}}) \vec{n}_{{\bf k}} + \frac{1}{2} \vec{V}_{\bot}({\bf k}).
\end{eqnarray}
in which $\vec{V}_{\bot}({\bf k})=\vec{V}({\bf k}) - [\hat{n}_{{\bf k}} \cdot \vec{V}({\bf k})] \hat{n}_{{\bf k}}$. Taking advantage of this effective Hamiltonian, we can use the winding number given in the previous section to analyze the topological properties of our system. It is found that, for the parameters of Fig.\ref{ES1}, the driving induces a topological phase transition from topological trivial superfluid to Floquet topological superfluid at $\omega=2E_{(\pi,\pi)}=20$, which is consistent with our previous statement.

%

Decreasing the driving frequency further, the gap at $\varepsilon_{0}$ can be closed and reopened again at $\omega=\{E_{{\bf k}_{c}}\}_{Max}$. The robust Floquet edge states will span both the gaps at $\varepsilon_{0}$ and $\varepsilon_{\pi}$ as shown in Fig.\ref{ES1}(d). These chiral edge states also localized at the boundaries of the system.

An interesting phenomenon is that both the $\varepsilon_{0,\pi}$ gaps are spanned by a pair chiral Floquet edge states, as shown in Fig.\ref{ES1}(d). Due to the bulk-boundary correspondence, the winding number of a band is equal to the difference between the number of edge states at the gaps above and below the band, in other words, we have  $C_{\epsilon\epsilon'}=n_{edge}(\epsilon)-n_{edge}(\epsilon')$. Therefore, the winding number of all the Floquet bands in Fig.\ref{ES1}(d) are zero. This is very different from the static cases, for which the robust chiral edge states only span the gap at $\varepsilon_{0}$, and only when the total winding number of bands below zero energy is nonzero. This anomalous bulk-edge correspondence also emerge in Floquet topological insulator \cite{MSRudnerPRX2013}.  As another prominent difference, the Floquet topological superfluid can possess two types robust Floquet edge states within the gaps at $\varepsilon_{0}$ and $\varepsilon_{\pi}$ respectively\cite{LJiangPRL2011}.  This will induce two types Floquet Majorana zero modes at topological defects of Floquet topological superfluid, which we will discuss in the next sections. Certainly, we can also create Floquet topological superfluid by adding suitable periodic driving to a topological nontrivial superfluid for $0<\mu< 8J_{0}$. There also exist two flavors of robust chiral edge states at the boundaries, even though the winding number of all the Foquet bands are zero.

In above analysis the topological phase transition does not depend on the driving strength $J_{0}$, which only relates the values of the inverse band gaps at $\varepsilon_{0,\pi}$.

\section{Phase Transition between Strong and Weak Pairing phases}
For $p_{x}+ip_{y}$ paired system\cite{NReadPRB2000}, the superfluid is topological trivial for strong pairing phase ($\mu<0$) and nontrivial for weak pairing phase ($\mu>0$). For the topological nontrivial phase, the pairing only changes the Fermi surface slightly. So, the Fermi surfaces is clear. Nevertheless, there is no Fermi surface in strong paired topological trivial phase. Now, we show that the above phase transition induced by periodic driving is transition from strong to weak pairing phase.

Firstly, we use the changing of the Fermi surface to show the transition between strong and weak pairing. In the driven superfluid system, we can get the density distribution from the Floquet equation(\ref{floqeq}). The Floquet state can be written as $\Psi({\bf k},t)=(u({\bf k},t),\upsilon({\bf k},t))^{T}$ and  $\phi_{m}({\bf k})=(u_{m}({\bf k}),\upsilon_{m}({\bf k}))^{T}$ the density distribution is
\begin{eqnarray}
n({\bf k})&=&\frac{1}{T}\int_{0}^{T} dt \upsilon({\bf k},t) \upsilon^{*}({\bf k},t)\nonumber\\
&=& \sum_{m} |\upsilon_{m}({\bf k})|^{2}.
\end{eqnarray}

\begin{figure}
\includegraphics[width=8.5cm, height=6.5cm]{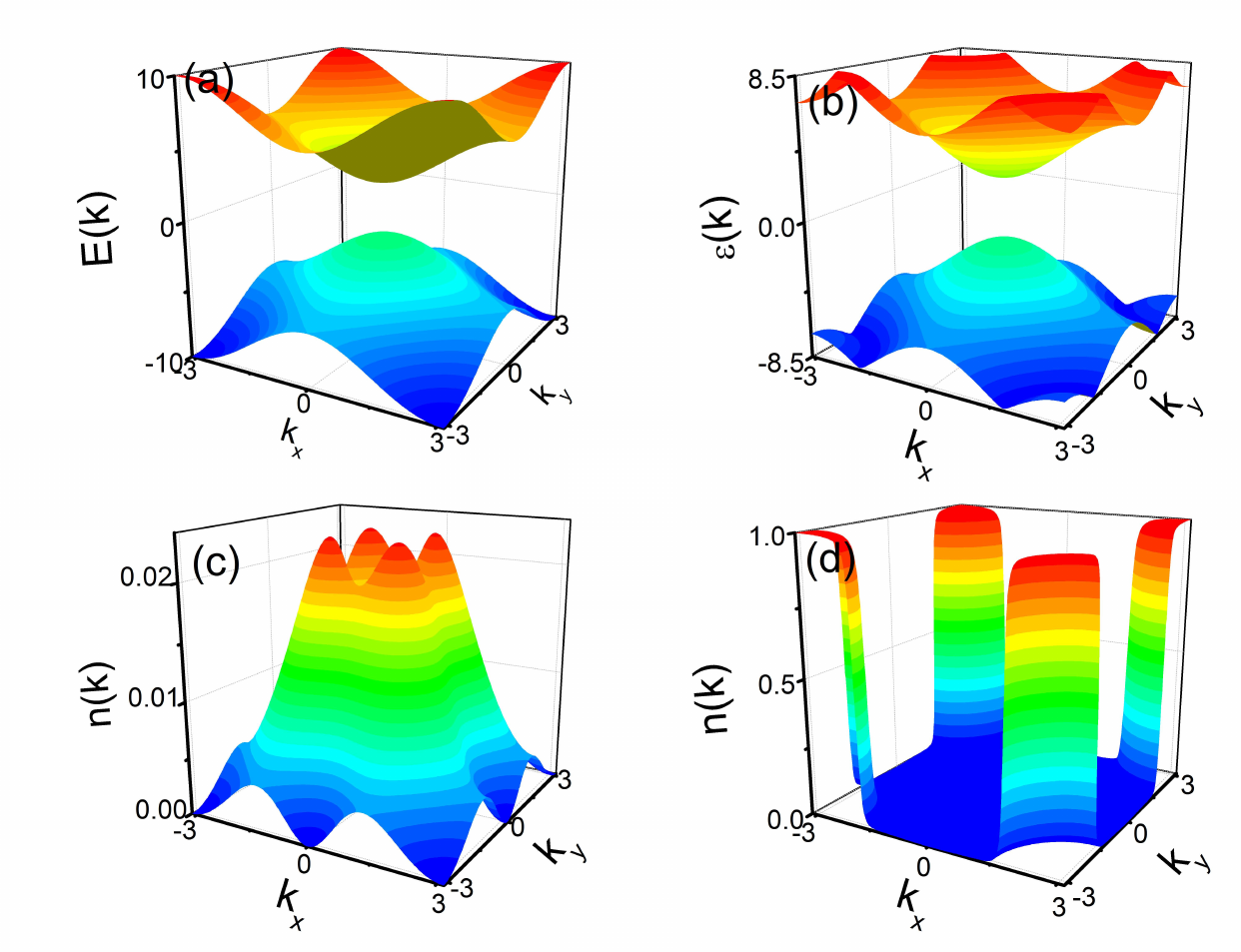}
\caption{(a) and (b) are the dispersion of the two superfluid phases with the parameters as Fig.\ref{ES1}(a) and (b) respectively. (c) and (d) are the density distribution at the momentum space of the (a) and (b). The density of ${\bf k}=(\pi,\pi)$ point has a jump from $0$ to $1$ as decreasing $\omega$. The jump is induced by a topological phase transition at $\omega=2 E_{{\bf k}=(\pi,\pi)}$.} \label{Dens}
\end{figure}

Fig.\ref{Dens} shows the dispersions and density distributions for the two cases of Fig.\ref{ES1}(a) and (b). The driving only changes the dispersion slightly as shown in Fig.\ref{Dens}(a)-(b) but completely changes the density distribution as shown in Fig.\ref{Dens}(c)-(d). For the topological trivial cases, the pairing is strong and there is no Fermi surface as shown in Fig.\ref{Dens}(c). In the presence of periodic driving, the Floquet topological superfluid has clear Fermi surfaces in the density distribution as shown in Fig.\ref{Dens}(d). Therefore, the periodic driving drives the strong paired topological trivial phase into weak paired Floquet topological phase.

Fig.\ref{Dens} also shows the density of TRI momentum $(\pi,\pi)$ has a universal jump at $\omega=2E_{(\pi,\pi)}$. The density of other TRI momenta will also have jump. The density will jump from $0$ to $1$ at $\omega=2E_{(0,\pi)}$ for $[(\pi,0),(0,\pi)]$ and $\omega=E_{(0,0)}$ for $(0,0)$. The jump is induced by the topological phase transition and can be used to distinguish the topological distinct phases.

Lastly, we use the effective Hamiltonian to analyze the topological phase transition at $\omega=2 E_{(\pi,\pi)}$. Here, we only consider the properties of system near the phase transition point and let $\omega=2 E_{(\pi,\pi)} + \delta$.  So, the effective Hamiltonian can be expanded at ${\bf k}_{c}=(\pi, \pi)$ and can be written as $\vec{n}'_{{\bf k}}=(A(\delta) k_{x}, A(\delta) k_{y}, -\delta \text{sgn}(\varepsilon_{k_{c}})/2)$ with $A(\delta)=\frac{\delta}{2E_{(\pi,\pi)}}+\frac{\varepsilon_{k_{c}}^{D}\varepsilon_{k_{c}}}{\varepsilon_{k_{c}}^{2}}$. Rewriting $\Delta'= A(\delta)$ and $\mu'=\delta \text{sgn}(\varepsilon_{k_{c}})/2$, the effective Hamiltonian has the same form with Ref.\cite{NReadPRB2000}. Thus, the superfluid is strong pairing phase for $\omega >2 E_{(\pi,\pi)}$, weak pairing phase for $\omega < 2 E_{(\pi,\pi)}$ and the transition is at $\omega=2 E_{(\pi,\pi)}$.

\section{Floquet Majorana Zero Modes}
\begin{figure}
\includegraphics[width=6.5cm, height=5cm]{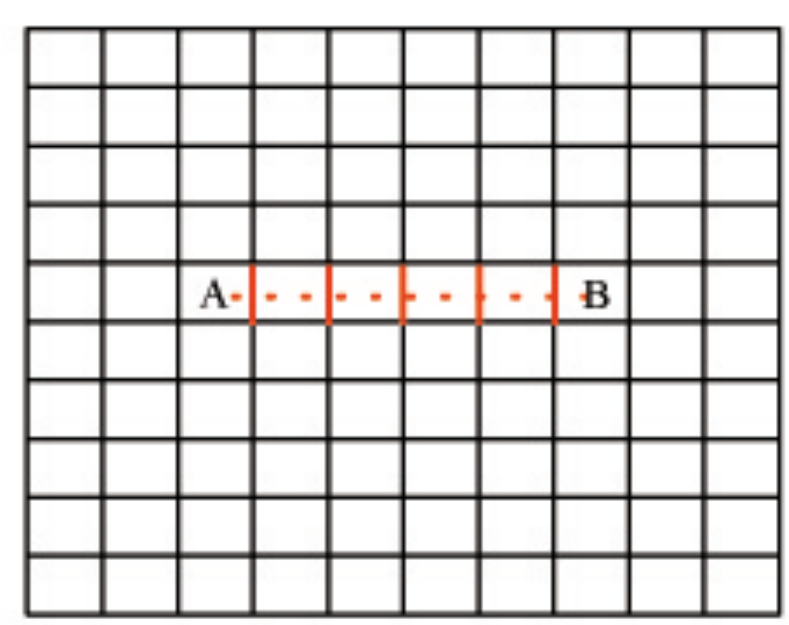}
\caption{The periodic lattice with two '$\pi$-Flux' at A and B sites.  The sign of all the links cut by dotted line will be changed.} \label{sys}
\end{figure}

Fundamentally different from the static topological superfluid phase, Floquet topological superfluid phase has nontrivial robust Floquet edge states spanning the gap at $\varepsilon_{\pi}$ \cite{LJiangPRL2011}. Now, we will show that this can be used to generate Floquet Majorana zero modes with finite quasienergy $\varepsilon_{\pi}$ at topological defects. To generate the Floquet Majorana zero modes, we add two '$\pi$-Flux' to the driven systems by changing the sign of the links cut by the line between two separate sites 'A' and 'B' as shown in Fig.\ref{sys}. Fermions will get a $\pi$ phase by circling 'A' or 'B' sites.

\begin{figure}
\includegraphics[width=8.5cm, height=6.5cm]{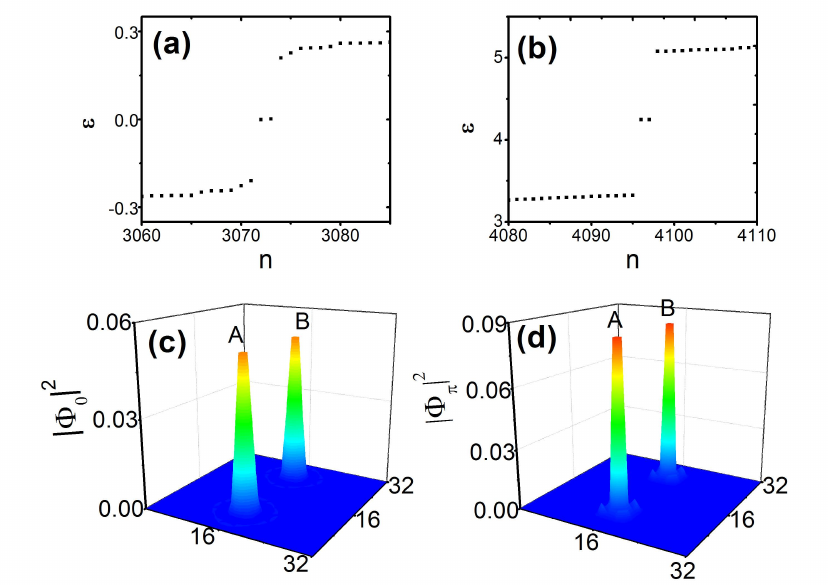}
\caption{(a) and (b) are the quasienergies around $\varepsilon_{0,\pi}$ respectively of the driven system with parameters as Fig.\ref{ES1}(d). The two pairs inner gap quasienergies $\varepsilon_{0,\pi}$ are the quasienergies the two types Floquet Majorana zero modes. (c) and (d) are the wave functions of two types Floquet Majorana zero modes ($|\Phi_{0,\pi}|^2$) of the system. The two types Floquet Majorana zero modes are localized at the 'flux' sites.} \label{FM}
\end{figure}

Fig.\ref{FM}(a) and (b) show the quasienergies, only around $\varepsilon_{0}$  and $\varepsilon_{\pi}$ respectively, of the topological defected system with the parameters as Fig.\ref{ES1}(d). For the particle-hole symmetry, there are two degenerate inner gap quasienergies $\varepsilon_{0}$ in Fig.\ref{FM}(a) and $\varepsilon_{\pi}$ in Fig.\ref{FM}(b). The wave functions of the two degenerate quasienergies also conjugate with each other. Fig.\ref{FM}(c) and (d) show the wave function with quasienergies $\varepsilon_{0}$ and $\varepsilon_{\pi}$ respectively. The wave functions $\Phi_{0}$ and $\Phi_{\pi}$ are localized at the same 'Flux' sites but separated in the quasienergy space at $\varepsilon_{0}$ and $\varepsilon_{\pi}$ respectively. These localized wave functions are the Floquet Majorana zero modes. Majorana zero modes are the states that they are their own conjugate. As for the particle-hole symmetry of the superfluid, the Majorana zero modes only can exist at $\varepsilon_{0}$ in static systems for only the state of $\varepsilon_{0}$ can be their own particle-hole conjugate. In the presence of periodic driving, the quasienergies of the driven systems are periodic. Therefore, the states of $\varepsilon_{\pi}=\pi/T\equiv-\pi/T$ also can be their own particle-hole conjugate. Thus, there exists two types Floquet Majorana zero modes with quasienergy $\varepsilon_{0,\pi}$ in Floquet topological superfluid. Here, the Floquet Majorana zero modes can be thought as the localization of the chiral Floquet edge states at the topological defects.

\section{Conclusion}
In the present paper, we have proposed a simple scheme to realize the two-dimensional Floquet topological nontrivial superfluid by periodically tuning the depth of square optical lattice potentials. The periodic driving can induces a transition from strong pairing phase to weak pairing phase. The weak pairing phases are Floquet topological superfluid phases and have clear Fermi surfaces. We have also found that there are two flavors Floquet Majorana zero modes at the topological defects of the Floquet topological superfluid phases, which may have potential applications in topological quantum computation.

\section{Acknowledge}
We especially grateful to Hai-Qing Lin and Zhong Wang for fruitful
discussion and collaborations. We also thank Wei Yi for helpful discussion. This
work is supported by NSFC 91230203, CAEP, and China
Postdoctoral Science Foundation (No. 2012M520147).

\bibliography{reference}

\begin{thebibliography}{45}%
\makeatletter
\providecommand \@ifxundefined [1]{%
 \@ifx{#1\undefined}
}%
\providecommand \@ifnum [1]{%
 \ifnum #1\expandafter \@firstoftwo
 \else \expandafter \@secondoftwo
 \fi
}%
\providecommand \@ifx [1]{%
 \ifx #1\expandafter \@firstoftwo
 \else \expandafter \@secondoftwo
 \fi
}%
\providecommand \natexlab [1]{#1}%
\providecommand \enquote  [1]{``#1''}%
\providecommand \bibnamefont  [1]{#1}%
\providecommand \bibfnamefont [1]{#1}%
\providecommand \citenamefont [1]{#1}%
\providecommand \href@noop [0]{\@secondoftwo}%
\providecommand \href [0]{\begingroup \@sanitize@url \@href}%
\providecommand \@href[1]{\@@startlink{#1}\@@href}%
\providecommand \@@href[1]{\endgroup#1\@@endlink}%
\providecommand \@sanitize@url [0]{\catcode `\\12\catcode `\$12\catcode
  `\&12\catcode `\#12\catcode `\^12\catcode `\_12\catcode `\%12\relax}%
\providecommand \@@startlink[1]{}%
\providecommand \@@endlink[0]{}%
\providecommand \url  [0]{\begingroup\@sanitize@url \@url }%
\providecommand \@url [1]{\endgroup\@href {#1}{\urlprefix }}%
\providecommand \urlprefix  [0]{URL }%
\providecommand \Eprint [0]{\href }%
\providecommand \doibase [0]{http://dx.doi.org/}%
\providecommand \selectlanguage [0]{\@gobble}%
\providecommand \bibinfo  [0]{\@secondoftwo}%
\providecommand \bibfield  [0]{\@secondoftwo}%
\providecommand \translation [1]{[#1]}%
\providecommand \BibitemOpen [0]{}%
\providecommand \bibitemStop [0]{}%
\providecommand \bibitemNoStop [0]{.\EOS\space}%
\providecommand \EOS [0]{\spacefactor3000\relax}%
\providecommand \BibitemShut  [1]{\csname bibitem#1\endcsname}%
\let\auto@bib@innerbib\@empty
\bibitem [{\citenamefont {Hasan}\ and\ \citenamefont
  {Kane}(2010)}]{HasanRevModPhys2010}%
  \BibitemOpen
  \bibfield  {author} {\bibinfo {author} {\bibfnamefont {M.~Z.}\ \bibnamefont
  {Hasan}}\ and\ \bibinfo {author} {\bibfnamefont {C.~L.}\ \bibnamefont
  {Kane}},\ }\href {\doibase 10.1103/RevModPhys.82.3045} {\bibfield  {journal}
  {\bibinfo  {journal} {Rev. Mod. Phys.}\ }\textbf {\bibinfo {volume} {82}},\
  \bibinfo {pages} {3045} (\bibinfo {year} {2010})}\BibitemShut {NoStop}%
\bibitem [{\citenamefont {Qi}\ and\ \citenamefont
  {Zhang}(2011)}]{XLQiRevModPhys2011}%
  \BibitemOpen
  \bibfield  {author} {\bibinfo {author} {\bibfnamefont {X.-L.}\ \bibnamefont
  {Qi}}\ and\ \bibinfo {author} {\bibfnamefont {S.-C.}\ \bibnamefont {Zhang}},\
  }\href {\doibase 10.1103/RevModPhys.83.1057} {\bibfield  {journal} {\bibinfo
  {journal} {Rev. Mod. Phys.}\ }\textbf {\bibinfo {volume} {83}},\ \bibinfo
  {pages} {1057} (\bibinfo {year} {2011})}\BibitemShut {NoStop}%
\bibitem [{\citenamefont {Thouless}\ \emph {et~al.}(1982)\citenamefont
  {Thouless}, \citenamefont {Kohmoto}, \citenamefont {Nightingale},\ and\
  \citenamefont {den Nijs}}]{thouless1982}%
  \BibitemOpen
  \bibfield  {author} {\bibinfo {author} {\bibfnamefont {D.~J.}\ \bibnamefont
  {Thouless}}, \bibinfo {author} {\bibfnamefont {M.}~\bibnamefont {Kohmoto}},
  \bibinfo {author} {\bibfnamefont {M.~P.}\ \bibnamefont {Nightingale}}, \ and\
  \bibinfo {author} {\bibfnamefont {M.}~\bibnamefont {den Nijs}},\ }\href@noop
  {} {\bibfield  {journal} {\bibinfo  {journal} {Phys. Rev. Lett.}\ }\textbf
  {\bibinfo {volume} {49}},\ \bibinfo {pages} {405} (\bibinfo {year}
  {1982})}\BibitemShut {NoStop}%
\bibitem [{\citenamefont {\textrm{C. L. Kane}}\ and\ \citenamefont {\textrm{E.
  J. Mele}}(2005)}]{kane2005b}%
  \BibitemOpen
  \bibfield  {author} {\bibinfo {author} {\bibnamefont {\textrm{C. L. Kane}}}\
  and\ \bibinfo {author} {\bibnamefont {\textrm{E. J. Mele}}},\ }\href@noop {}
  {\bibfield  {journal} {\bibinfo  {journal} {Phys. Rev. Lett.}\ }\textbf
  {\bibinfo {volume} {95}},\ \bibinfo {pages} {146802} (\bibinfo {year}
  {2005})}\BibitemShut {NoStop}%
\bibitem [{\citenamefont {Fu}\ \emph {et~al.}(2007)\citenamefont {Fu},
  \citenamefont {Kane},\ and\ \citenamefont {Mele}}]{fu2007b}%
  \BibitemOpen
  \bibfield  {author} {\bibinfo {author} {\bibfnamefont {L.}~\bibnamefont
  {Fu}}, \bibinfo {author} {\bibfnamefont {C.~L.}\ \bibnamefont {Kane}}, \ and\
  \bibinfo {author} {\bibfnamefont {E.~J.}\ \bibnamefont {Mele}},\ }\href
  {\doibase 10.1103/PhysRevLett.98.106803} {\bibfield  {journal} {\bibinfo
  {journal} {Phys. Rev. Lett.}\ }\textbf {\bibinfo {volume} {98}},\ \bibinfo
  {eid} {106803} (\bibinfo {year} {2007})}\BibitemShut {NoStop}%
\bibitem [{\citenamefont {Qi}\ \emph {et~al.}(2008)\citenamefont {Qi},
  \citenamefont {Hughes},\ and\ \citenamefont {Zhang}}]{qi2008}%
  \BibitemOpen
  \bibfield  {author} {\bibinfo {author} {\bibfnamefont {X.-L.}\ \bibnamefont
  {Qi}}, \bibinfo {author} {\bibfnamefont {T.}~\bibnamefont {Hughes}}, \ and\
  \bibinfo {author} {\bibfnamefont {S.-C.}\ \bibnamefont {Zhang}},\ }\href@noop
  {} {\bibfield  {journal} {\bibinfo  {journal} {Phys. Rev. B}\ }\textbf
  {\bibinfo {volume} {78}},\ \bibinfo {pages} {195424} (\bibinfo {year}
  {2008})}\BibitemShut {NoStop}%
\bibitem [{\citenamefont {Wang}\ \emph
  {et~al.}(2010{\natexlab{a}})\citenamefont {Wang}, \citenamefont {Qi},\ and\
  \citenamefont {Zhang}}]{wang2010order}%
  \BibitemOpen
  \bibfield  {author} {\bibinfo {author} {\bibfnamefont {Z.}~\bibnamefont
  {Wang}}, \bibinfo {author} {\bibfnamefont {X.-L.}\ \bibnamefont {Qi}}, \ and\
  \bibinfo {author} {\bibfnamefont {S.-C.}\ \bibnamefont {Zhang}},\ }\href@noop
  {} {\bibfield  {journal} {\bibinfo  {journal} {Phys. Rev. Lett.}\ }\textbf
  {\bibinfo {volume} {105}},\ \bibinfo {pages} {256803} (\bibinfo {year}
  {2010}{\natexlab{a}})}\BibitemShut {NoStop}%
\bibitem [{\citenamefont {Wang}\ \emph
  {et~al.}(2010{\natexlab{b}})\citenamefont {Wang}, \citenamefont {Qi},\ and\
  \citenamefont {Zhang}}]{wang2010a}%
  \BibitemOpen
  \bibfield  {author} {\bibinfo {author} {\bibfnamefont {Z.}~\bibnamefont
  {Wang}}, \bibinfo {author} {\bibfnamefont {X.-L.}\ \bibnamefont {Qi}}, \ and\
  \bibinfo {author} {\bibfnamefont {S.-C.}\ \bibnamefont {Zhang}},\ }\href@noop
  {} {\bibfield  {journal} {\bibinfo  {journal} {New J. Phys.}\ }\textbf
  {\bibinfo {volume} {12}},\ \bibinfo {pages} {065007} (\bibinfo {year}
  {2010}{\natexlab{b}})}\BibitemShut {NoStop}%
\bibitem [{\citenamefont {Wang}\ and\ \citenamefont
  {Zhang}(2012)}]{wang2012simplified}%
  \BibitemOpen
  \bibfield  {author} {\bibinfo {author} {\bibfnamefont {Z.}~\bibnamefont
  {Wang}}\ and\ \bibinfo {author} {\bibfnamefont {S.-C.}\ \bibnamefont
  {Zhang}},\ }\href {\doibase 10.1103/PhysRevX.2.031008} {\bibfield  {journal}
  {\bibinfo  {journal} {Phys. Rev. X}\ }\textbf {\bibinfo {volume} {2}},\
  \bibinfo {pages} {031008} (\bibinfo {year} {2012})}\BibitemShut {NoStop}%
\bibitem [{\citenamefont {Eckardt}\ \emph {et~al.}(2005)\citenamefont
  {Eckardt}, \citenamefont {Weiss},\ and\ \citenamefont
  {Holthaus}}]{eckardtPhysRevLett2005}%
  \BibitemOpen
  \bibfield  {author} {\bibinfo {author} {\bibfnamefont {A.}~\bibnamefont
  {Eckardt}}, \bibinfo {author} {\bibfnamefont {C.}~\bibnamefont {Weiss}}, \
  and\ \bibinfo {author} {\bibfnamefont {M.}~\bibnamefont {Holthaus}},\ }\href
  {\doibase 10.1103/PhysRevLett.95.260404} {\bibfield  {journal} {\bibinfo
  {journal} {Phys. Rev. Lett.}\ }\textbf {\bibinfo {volume} {95}},\ \bibinfo
  {pages} {260404} (\bibinfo {year} {2005})}\BibitemShut {NoStop}%
\bibitem [{\citenamefont {Kitagawa}\ \emph {et~al.}(2010)\citenamefont
  {Kitagawa}, \citenamefont {Berg}, \citenamefont {Rudner},\ and\ \citenamefont
  {Demler}}]{TKitagawaPRB2010}%
  \BibitemOpen
  \bibfield  {author} {\bibinfo {author} {\bibfnamefont {T.}~\bibnamefont
  {Kitagawa}}, \bibinfo {author} {\bibfnamefont {E.}~\bibnamefont {Berg}},
  \bibinfo {author} {\bibfnamefont {M.}~\bibnamefont {Rudner}}, \ and\ \bibinfo
  {author} {\bibfnamefont {E.}~\bibnamefont {Demler}},\ }\href {\doibase
  10.1103/PhysRevB.82.235114} {\bibfield  {journal} {\bibinfo  {journal} {Phys.
  Rev. B}\ }\textbf {\bibinfo {volume} {82}},\ \bibinfo {pages} {235114}
  (\bibinfo {year} {2010})}\BibitemShut {NoStop}%
\bibitem [{\citenamefont {Lindner}\ \emph {et~al.}(2011)\citenamefont
  {Lindner}, \citenamefont {Refael},\ and\ \citenamefont
  {Galitski}}]{NLindnerNATPHY2011}%
  \BibitemOpen
  \bibfield  {author} {\bibinfo {author} {\bibfnamefont {N.~H.}\ \bibnamefont
  {Lindner}}, \bibinfo {author} {\bibfnamefont {G.}~\bibnamefont {Refael}}, \
  and\ \bibinfo {author} {\bibfnamefont {V.}~\bibnamefont {Galitski}},\
  }\href@noop {} {\bibfield  {journal} {\bibinfo  {journal} {Nature Physics}\
  }\textbf {\bibinfo {volume} {7}},\ \bibinfo {pages} {490} (\bibinfo {year}
  {2011})}\BibitemShut {NoStop}%
\bibitem [{\citenamefont {Kundu}\ and\ \citenamefont
  {Seradjeh}(2013)}]{KArijitPRL2013}%
  \BibitemOpen
  \bibfield  {author} {\bibinfo {author} {\bibfnamefont {A.}~\bibnamefont
  {Kundu}}\ and\ \bibinfo {author} {\bibfnamefont {B.}~\bibnamefont
  {Seradjeh}},\ }\href {\doibase 10.1103/PhysRevLett.111.136402} {\bibfield
  {journal} {\bibinfo  {journal} {Phys. Rev. Lett.}\ }\textbf {\bibinfo
  {volume} {111}},\ \bibinfo {pages} {136402} (\bibinfo {year}
  {2013})}\BibitemShut {NoStop}%
\bibitem [{\citenamefont {Liu}\ \emph {et~al.}(2013)\citenamefont {Liu},
  \citenamefont {Levchenko},\ and\ \citenamefont {Baranger}}]{DEliuPRL2013}%
  \BibitemOpen
  \bibfield  {author} {\bibinfo {author} {\bibfnamefont {D.~E.}\ \bibnamefont
  {Liu}}, \bibinfo {author} {\bibfnamefont {A.}~\bibnamefont {Levchenko}}, \
  and\ \bibinfo {author} {\bibfnamefont {H.~U.}\ \bibnamefont {Baranger}},\
  }\href {\doibase 10.1103/PhysRevLett.111.047002} {\bibfield  {journal}
  {\bibinfo  {journal} {Phys. Rev. Lett.}\ }\textbf {\bibinfo {volume} {111}},\
  \bibinfo {pages} {047002} (\bibinfo {year} {2013})}\BibitemShut {NoStop}%
\bibitem [{\citenamefont {Jiang}\ \emph {et~al.}(2011)\citenamefont {Jiang},
  \citenamefont {Kitagawa}, \citenamefont {Alicea}, \citenamefont {Akhmerov},
  \citenamefont {Pekker}, \citenamefont {Refael}, \citenamefont {Cirac},
  \citenamefont {Demler}, \citenamefont {Lukin},\ and\ \citenamefont
  {Zoller}}]{LJiangPRL2011}%
  \BibitemOpen
  \bibfield  {author} {\bibinfo {author} {\bibfnamefont {L.}~\bibnamefont
  {Jiang}}, \bibinfo {author} {\bibfnamefont {T.}~\bibnamefont {Kitagawa}},
  \bibinfo {author} {\bibfnamefont {J.}~\bibnamefont {Alicea}}, \bibinfo
  {author} {\bibfnamefont {A.~R.}\ \bibnamefont {Akhmerov}}, \bibinfo {author}
  {\bibfnamefont {D.}~\bibnamefont {Pekker}}, \bibinfo {author} {\bibfnamefont
  {G.}~\bibnamefont {Refael}}, \bibinfo {author} {\bibfnamefont {J.~I.}\
  \bibnamefont {Cirac}}, \bibinfo {author} {\bibfnamefont {E.}~\bibnamefont
  {Demler}}, \bibinfo {author} {\bibfnamefont {M.~D.}\ \bibnamefont {Lukin}}, \
  and\ \bibinfo {author} {\bibfnamefont {P.}~\bibnamefont {Zoller}},\ }\href
  {\doibase 10.1103/PhysRevLett.106.220402} {\bibfield  {journal} {\bibinfo
  {journal} {Phys. Rev. Lett.}\ }\textbf {\bibinfo {volume} {106}},\ \bibinfo
  {pages} {220402} (\bibinfo {year} {2011})}\BibitemShut {NoStop}%
\bibitem [{\citenamefont {Katan}\ and\ \citenamefont
  {Podolsky}(2013)}]{YKatanPRL2013}%
  \BibitemOpen
  \bibfield  {author} {\bibinfo {author} {\bibfnamefont {Y.~T.}\ \bibnamefont
  {Katan}}\ and\ \bibinfo {author} {\bibfnamefont {D.}~\bibnamefont
  {Podolsky}},\ }\href {\doibase 10.1103/PhysRevLett.110.016802} {\bibfield
  {journal} {\bibinfo  {journal} {Phys. Rev. Lett.}\ }\textbf {\bibinfo
  {volume} {110}},\ \bibinfo {pages} {016802} (\bibinfo {year}
  {2013})}\BibitemShut {NoStop}%
\bibitem [{\citenamefont {Wang}\ \emph {et~al.}(2013)\citenamefont {Wang},
  \citenamefont {Steinberg}, \citenamefont {Jarillo-Herrero},\ and\
  \citenamefont {Gedik}}]{YHWangScience2013}%
  \BibitemOpen
  \bibfield  {author} {\bibinfo {author} {\bibfnamefont {Y.}~\bibnamefont
  {Wang}}, \bibinfo {author} {\bibfnamefont {H.}~\bibnamefont {Steinberg}},
  \bibinfo {author} {\bibfnamefont {P.}~\bibnamefont {Jarillo-Herrero}}, \ and\
  \bibinfo {author} {\bibfnamefont {N.}~\bibnamefont {Gedik}},\ }\href@noop {}
  {\bibfield  {journal} {\bibinfo  {journal} {Science}\ }\textbf {\bibinfo
  {volume} {342}},\ \bibinfo {pages} {453} (\bibinfo {year}
  {2013})}\BibitemShut {NoStop}%
\bibitem [{\citenamefont {G\'omez-Le\'on}\ and\ \citenamefont
  {Platero}(2013)}]{LGomezPRL2013}%
  \BibitemOpen
  \bibfield  {author} {\bibinfo {author} {\bibfnamefont {A.}~\bibnamefont
  {G\'omez-Le\'on}}\ and\ \bibinfo {author} {\bibfnamefont {G.}~\bibnamefont
  {Platero}},\ }\href {\doibase 10.1103/PhysRevLett.110.200403} {\bibfield
  {journal} {\bibinfo  {journal} {Phys. Rev. Lett.}\ }\textbf {\bibinfo
  {volume} {110}},\ \bibinfo {pages} {200403} (\bibinfo {year}
  {2013})}\BibitemShut {NoStop}%
\bibitem [{\citenamefont {Rechtsman}\ \emph
  {et~al.}(2013{\natexlab{a}})\citenamefont {Rechtsman}, \citenamefont
  {Zeuner}, \citenamefont {Plotnik}, \citenamefont {Lumer}, \citenamefont
  {Podolsky}, \citenamefont {Dreisow}, \citenamefont {Nolte}, \citenamefont
  {Segev},\ and\ \citenamefont {Szameit}}]{rechtsmannature2013}%
  \BibitemOpen
  \bibfield  {author} {\bibinfo {author} {\bibfnamefont {M.~C.}\ \bibnamefont
  {Rechtsman}}, \bibinfo {author} {\bibfnamefont {J.~M.}\ \bibnamefont
  {Zeuner}}, \bibinfo {author} {\bibfnamefont {Y.}~\bibnamefont {Plotnik}},
  \bibinfo {author} {\bibfnamefont {Y.}~\bibnamefont {Lumer}}, \bibinfo
  {author} {\bibfnamefont {D.}~\bibnamefont {Podolsky}}, \bibinfo {author}
  {\bibfnamefont {F.}~\bibnamefont {Dreisow}}, \bibinfo {author} {\bibfnamefont
  {S.}~\bibnamefont {Nolte}}, \bibinfo {author} {\bibfnamefont
  {M.}~\bibnamefont {Segev}}, \ and\ \bibinfo {author} {\bibfnamefont
  {A.}~\bibnamefont {Szameit}},\ }\href@noop {} {\bibfield  {journal} {\bibinfo
   {journal} {Nature}\ }\textbf {\bibinfo {volume} {496}},\ \bibinfo {pages}
  {196} (\bibinfo {year} {2013}{\natexlab{a}})}\BibitemShut {NoStop}%
\bibitem [{\citenamefont {Lababidi}\ \emph {et~al.}(2014)\citenamefont
  {Lababidi}, \citenamefont {Satija},\ and\ \citenamefont
  {Zhao}}]{MLababidiPRL2014}%
  \BibitemOpen
  \bibfield  {author} {\bibinfo {author} {\bibfnamefont {M.}~\bibnamefont
  {Lababidi}}, \bibinfo {author} {\bibfnamefont {I.~I.}\ \bibnamefont
  {Satija}}, \ and\ \bibinfo {author} {\bibfnamefont {E.}~\bibnamefont
  {Zhao}},\ }\href {\doibase 10.1103/PhysRevLett.112.026805} {\bibfield
  {journal} {\bibinfo  {journal} {Phys. Rev. Lett.}\ }\textbf {\bibinfo
  {volume} {112}},\ \bibinfo {pages} {026805} (\bibinfo {year}
  {2014})}\BibitemShut {NoStop}%
\bibitem [{\citenamefont {Sato}\ \emph {et~al.}(2014)\citenamefont {Sato},
  \citenamefont {Sasaki},\ and\ \citenamefont {Oka}}]{satoarxiv2014}%
  \BibitemOpen
  \bibfield  {author} {\bibinfo {author} {\bibfnamefont {M.}~\bibnamefont
  {Sato}}, \bibinfo {author} {\bibfnamefont {Y.}~\bibnamefont {Sasaki}}, \ and\
  \bibinfo {author} {\bibfnamefont {T.}~\bibnamefont {Oka}},\ }\href@noop {}
  {\bibfield  {journal} {\bibinfo  {journal} {arXiv preprint arXiv:1404.2010}\
  } (\bibinfo {year} {2014})}\BibitemShut {NoStop}%
\bibitem [{\citenamefont {Rudner}\ \emph {et~al.}(2013)\citenamefont {Rudner},
  \citenamefont {Lindner}, \citenamefont {Berg},\ and\ \citenamefont
  {Levin}}]{MSRudnerPRX2013}%
  \BibitemOpen
  \bibfield  {author} {\bibinfo {author} {\bibfnamefont {M.~S.}\ \bibnamefont
  {Rudner}}, \bibinfo {author} {\bibfnamefont {N.~H.}\ \bibnamefont {Lindner}},
  \bibinfo {author} {\bibfnamefont {E.}~\bibnamefont {Berg}}, \ and\ \bibinfo
  {author} {\bibfnamefont {M.}~\bibnamefont {Levin}},\ }\href {\doibase
  10.1103/PhysRevX.3.031005} {\bibfield  {journal} {\bibinfo  {journal} {Phys.
  Rev. X}\ }\textbf {\bibinfo {volume} {3}},\ \bibinfo {pages} {031005}
  (\bibinfo {year} {2013})}\BibitemShut {NoStop}%
\bibitem [{\citenamefont {Lindner}\ \emph {et~al.}(2013)\citenamefont
  {Lindner}, \citenamefont {Bergman}, \citenamefont {Refael},\ and\
  \citenamefont {Galitski}}]{lindnerPhysRevB2013}%
  \BibitemOpen
  \bibfield  {author} {\bibinfo {author} {\bibfnamefont {N.~H.}\ \bibnamefont
  {Lindner}}, \bibinfo {author} {\bibfnamefont {D.~L.}\ \bibnamefont
  {Bergman}}, \bibinfo {author} {\bibfnamefont {G.}~\bibnamefont {Refael}}, \
  and\ \bibinfo {author} {\bibfnamefont {V.}~\bibnamefont {Galitski}},\ }\href
  {\doibase 10.1103/PhysRevB.87.235131} {\bibfield  {journal} {\bibinfo
  {journal} {Phys. Rev. B}\ }\textbf {\bibinfo {volume} {87}},\ \bibinfo
  {pages} {235131} (\bibinfo {year} {2013})}\BibitemShut {NoStop}%
\bibitem [{\citenamefont {Rechtsman}\ \emph
  {et~al.}(2013{\natexlab{b}})\citenamefont {Rechtsman}, \citenamefont
  {Zeuner}, \citenamefont {Plotnik}, \citenamefont {Lumer}, \citenamefont
  {Podolsky}, \citenamefont {Dreisow}, \citenamefont {Nolte}, \citenamefont
  {Segev},\ and\ \citenamefont {Szameit}}]{MRechtsmanNature2013}%
  \BibitemOpen
  \bibfield  {author} {\bibinfo {author} {\bibfnamefont {M.~C.}\ \bibnamefont
  {Rechtsman}}, \bibinfo {author} {\bibfnamefont {J.~M.}\ \bibnamefont
  {Zeuner}}, \bibinfo {author} {\bibfnamefont {Y.}~\bibnamefont {Plotnik}},
  \bibinfo {author} {\bibfnamefont {Y.}~\bibnamefont {Lumer}}, \bibinfo
  {author} {\bibfnamefont {D.}~\bibnamefont {Podolsky}}, \bibinfo {author}
  {\bibfnamefont {F.}~\bibnamefont {Dreisow}}, \bibinfo {author} {\bibfnamefont
  {S.}~\bibnamefont {Nolte}}, \bibinfo {author} {\bibfnamefont
  {M.}~\bibnamefont {Segev}}, \ and\ \bibinfo {author} {\bibfnamefont
  {A.}~\bibnamefont {Szameit}},\ }\href@noop {} {\bibfield  {journal} {\bibinfo
   {journal} {Nature}\ }\textbf {\bibinfo {volume} {496}},\ \bibinfo {pages}
  {196} (\bibinfo {year} {2013}{\natexlab{b}})}\BibitemShut {NoStop}%
\bibitem [{\citenamefont {Tong}\ \emph {et~al.}(2013)\citenamefont {Tong},
  \citenamefont {An}, \citenamefont {Gong}, \citenamefont {Luo},\ and\
  \citenamefont {Oh}}]{QJTongPRL2013}%
  \BibitemOpen
  \bibfield  {author} {\bibinfo {author} {\bibfnamefont {Q.-J.}\ \bibnamefont
  {Tong}}, \bibinfo {author} {\bibfnamefont {J.-H.}\ \bibnamefont {An}},
  \bibinfo {author} {\bibfnamefont {J.}~\bibnamefont {Gong}}, \bibinfo {author}
  {\bibfnamefont {H.-G.}\ \bibnamefont {Luo}}, \ and\ \bibinfo {author}
  {\bibfnamefont {C.~H.}\ \bibnamefont {Oh}},\ }\href {\doibase
  10.1103/PhysRevB.87.201109} {\bibfield  {journal} {\bibinfo  {journal} {Phys.
  Rev. B}\ }\textbf {\bibinfo {volume} {87}},\ \bibinfo {pages} {201109}
  (\bibinfo {year} {2013})}\BibitemShut {NoStop}%
\bibitem [{\citenamefont {Grushin}\ \emph {et~al.}(2014)\citenamefont
  {Grushin}, \citenamefont {G\'omez-Le\'on},\ and\ \citenamefont
  {Neupert}}]{grushinPhysRevLett2014}%
  \BibitemOpen
  \bibfield  {author} {\bibinfo {author} {\bibfnamefont {A.~G.}\ \bibnamefont
  {Grushin}}, \bibinfo {author} {\bibfnamefont {A.}~\bibnamefont
  {G\'omez-Le\'on}}, \ and\ \bibinfo {author} {\bibfnamefont {T.}~\bibnamefont
  {Neupert}},\ }\href {\doibase 10.1103/PhysRevLett.112.156801} {\bibfield
  {journal} {\bibinfo  {journal} {Phys. Rev. Lett.}\ }\textbf {\bibinfo
  {volume} {112}},\ \bibinfo {pages} {156801} (\bibinfo {year}
  {2014})}\BibitemShut {NoStop}%
\bibitem [{\citenamefont {Wang}\ \emph {et~al.}(2014)\citenamefont {Wang},
  \citenamefont {Wang}, \citenamefont {Shen}, \citenamefont {Sheng},\ and\
  \citenamefont {Xing}}]{wangarxiv2014}%
  \BibitemOpen
  \bibfield  {author} {\bibinfo {author} {\bibfnamefont {R.}~\bibnamefont
  {Wang}}, \bibinfo {author} {\bibfnamefont {B.}~\bibnamefont {Wang}}, \bibinfo
  {author} {\bibfnamefont {R.}~\bibnamefont {Shen}}, \bibinfo {author}
  {\bibfnamefont {L.}~\bibnamefont {Sheng}}, \ and\ \bibinfo {author}
  {\bibfnamefont {D.}~\bibnamefont {Xing}},\ }\href@noop {} {\bibfield
  {journal} {\bibinfo  {journal} {EPL (Europhysics Letters)}\ }\textbf
  {\bibinfo {volume} {105}},\ \bibinfo {pages} {17004} (\bibinfo {year}
  {2014})}\BibitemShut {NoStop}%
\bibitem [{\citenamefont {G{\'o}mez-Le{\'o}n}\ \emph
  {et~al.}(2013)\citenamefont {G{\'o}mez-Le{\'o}n}, \citenamefont {Delplace},\
  and\ \citenamefont {Platero}}]{gomezarxiv2013}%
  \BibitemOpen
  \bibfield  {author} {\bibinfo {author} {\bibfnamefont {{\'A}.}~\bibnamefont
  {G{\'o}mez-Le{\'o}n}}, \bibinfo {author} {\bibfnamefont {P.}~\bibnamefont
  {Delplace}}, \ and\ \bibinfo {author} {\bibfnamefont {G.}~\bibnamefont
  {Platero}},\ }\href@noop {} {\bibfield  {journal} {\bibinfo  {journal} {arXiv
  preprint arXiv:1309.5402}\ } (\bibinfo {year} {2013})}\BibitemShut {NoStop}%
\bibitem [{\citenamefont {Titum}\ \emph {et~al.}(2014)\citenamefont {Titum},
  \citenamefont {Lindner}, \citenamefont {Rechtsman},\ and\ \citenamefont
  {Refael}}]{Titumdarxiv2014}%
  \BibitemOpen
  \bibfield  {author} {\bibinfo {author} {\bibfnamefont {P.}~\bibnamefont
  {Titum}}, \bibinfo {author} {\bibfnamefont {N.~H.}\ \bibnamefont {Lindner}},
  \bibinfo {author} {\bibfnamefont {M.~C.}\ \bibnamefont {Rechtsman}}, \ and\
  \bibinfo {author} {\bibfnamefont {G.}~\bibnamefont {Refael}},\ }\href@noop {}
  {\bibfield  {journal} {\bibinfo  {journal} {arXiv preprint arXiv:1403.0592}\
  } (\bibinfo {year} {2014})}\BibitemShut {NoStop}%
\bibitem [{\citenamefont {Lim}\ \emph {et~al.}(2008)\citenamefont {Lim},
  \citenamefont {Smith},\ and\ \citenamefont {Hemmerich}}]{limPhysRevLett2008}%
  \BibitemOpen
  \bibfield  {author} {\bibinfo {author} {\bibfnamefont {L.-K.}\ \bibnamefont
  {Lim}}, \bibinfo {author} {\bibfnamefont {C.~M.}\ \bibnamefont {Smith}}, \
  and\ \bibinfo {author} {\bibfnamefont {A.}~\bibnamefont {Hemmerich}},\ }\href
  {\doibase 10.1103/PhysRevLett.100.130402} {\bibfield  {journal} {\bibinfo
  {journal} {Phys. Rev. Lett.}\ }\textbf {\bibinfo {volume} {100}},\ \bibinfo
  {pages} {130402} (\bibinfo {year} {2008})}\BibitemShut {NoStop}%
\bibitem [{\citenamefont {Lim}\ \emph {et~al.}(2010)\citenamefont {Lim},
  \citenamefont {Hemmerich},\ and\ \citenamefont {Smith}}]{limPhysRevA2010}%
  \BibitemOpen
  \bibfield  {author} {\bibinfo {author} {\bibfnamefont {L.-K.}\ \bibnamefont
  {Lim}}, \bibinfo {author} {\bibfnamefont {A.}~\bibnamefont {Hemmerich}}, \
  and\ \bibinfo {author} {\bibfnamefont {C.~M.}\ \bibnamefont {Smith}},\ }\href
  {\doibase 10.1103/PhysRevA.81.023404} {\bibfield  {journal} {\bibinfo
  {journal} {Phys. Rev. A}\ }\textbf {\bibinfo {volume} {81}},\ \bibinfo
  {pages} {023404} (\bibinfo {year} {2010})}\BibitemShut {NoStop}%
\bibitem [{\citenamefont {Di~Liberto}\ \emph {et~al.}(2011)\citenamefont
  {Di~Liberto}, \citenamefont {Tieleman}, \citenamefont {Branchina},\ and\
  \citenamefont {Smith}}]{libertoPhysRevA2011}%
  \BibitemOpen
  \bibfield  {author} {\bibinfo {author} {\bibfnamefont {M.}~\bibnamefont
  {Di~Liberto}}, \bibinfo {author} {\bibfnamefont {O.}~\bibnamefont
  {Tieleman}}, \bibinfo {author} {\bibfnamefont {V.}~\bibnamefont {Branchina}},
  \ and\ \bibinfo {author} {\bibfnamefont {C.~M.}\ \bibnamefont {Smith}},\
  }\href {\doibase 10.1103/PhysRevA.84.013607} {\bibfield  {journal} {\bibinfo
  {journal} {Phys. Rev. A}\ }\textbf {\bibinfo {volume} {84}},\ \bibinfo
  {pages} {013607} (\bibinfo {year} {2011})}\BibitemShut {NoStop}%
\bibitem [{\citenamefont {Koghee}\ \emph {et~al.}(2012)\citenamefont {Koghee},
  \citenamefont {Lim}, \citenamefont {Goerbig},\ and\ \citenamefont
  {Smith}}]{kogheePhysRevA2012}%
  \BibitemOpen
  \bibfield  {author} {\bibinfo {author} {\bibfnamefont {S.}~\bibnamefont
  {Koghee}}, \bibinfo {author} {\bibfnamefont {L.-K.}\ \bibnamefont {Lim}},
  \bibinfo {author} {\bibfnamefont {M.~O.}\ \bibnamefont {Goerbig}}, \ and\
  \bibinfo {author} {\bibfnamefont {C.~M.}\ \bibnamefont {Smith}},\ }\href
  {\doibase 10.1103/PhysRevA.85.023637} {\bibfield  {journal} {\bibinfo
  {journal} {Phys. Rev. A}\ }\textbf {\bibinfo {volume} {85}},\ \bibinfo
  {pages} {023637} (\bibinfo {year} {2012})}\BibitemShut {NoStop}%
\bibitem [{\citenamefont {Hauke}\ \emph {et~al.}(2012)\citenamefont {Hauke},
  \citenamefont {Tieleman}, \citenamefont {Celi}, \citenamefont
  {\"Olschl\"ager}, \citenamefont {Simonet}, \citenamefont {Struck},
  \citenamefont {Weinberg}, \citenamefont {Windpassinger}, \citenamefont
  {Sengstock}, \citenamefont {Lewenstein},\ and\ \citenamefont
  {Eckardt}}]{haukePhysRevLett2012}%
  \BibitemOpen
  \bibfield  {author} {\bibinfo {author} {\bibfnamefont {P.}~\bibnamefont
  {Hauke}}, \bibinfo {author} {\bibfnamefont {O.}~\bibnamefont {Tieleman}},
  \bibinfo {author} {\bibfnamefont {A.}~\bibnamefont {Celi}}, \bibinfo {author}
  {\bibfnamefont {C.}~\bibnamefont {\"Olschl\"ager}}, \bibinfo {author}
  {\bibfnamefont {J.}~\bibnamefont {Simonet}}, \bibinfo {author} {\bibfnamefont
  {J.}~\bibnamefont {Struck}}, \bibinfo {author} {\bibfnamefont
  {M.}~\bibnamefont {Weinberg}}, \bibinfo {author} {\bibfnamefont
  {P.}~\bibnamefont {Windpassinger}}, \bibinfo {author} {\bibfnamefont
  {K.}~\bibnamefont {Sengstock}}, \bibinfo {author} {\bibfnamefont
  {M.}~\bibnamefont {Lewenstein}}, \ and\ \bibinfo {author} {\bibfnamefont
  {A.}~\bibnamefont {Eckardt}},\ }\href {\doibase
  10.1103/PhysRevLett.109.145301} {\bibfield  {journal} {\bibinfo  {journal}
  {Phys. Rev. Lett.}\ }\textbf {\bibinfo {volume} {109}},\ \bibinfo {pages}
  {145301} (\bibinfo {year} {2012})}\BibitemShut {NoStop}%
\bibitem [{\citenamefont {Zheng}\ and\ \citenamefont
  {Zhai}(2014)}]{zhengarxiv2014}%
  \BibitemOpen
  \bibfield  {author} {\bibinfo {author} {\bibfnamefont {W.}~\bibnamefont
  {Zheng}}\ and\ \bibinfo {author} {\bibfnamefont {H.}~\bibnamefont {Zhai}},\
  }\href@noop {} {\bibfield  {journal} {\bibinfo  {journal} {arXiv preprint
  arXiv:1402.4034}\ } (\bibinfo {year} {2014})}\BibitemShut {NoStop}%
\bibitem [{\citenamefont {Peil}\ \emph {et~al.}(2003)\citenamefont {Peil},
  \citenamefont {Porto}, \citenamefont {Tolra}, \citenamefont {Obrecht},
  \citenamefont {King}, \citenamefont {Subbotin}, \citenamefont {Rolston},\
  and\ \citenamefont {Phillips}}]{peilPhysRevA2003}%
  \BibitemOpen
  \bibfield  {author} {\bibinfo {author} {\bibfnamefont {S.}~\bibnamefont
  {Peil}}, \bibinfo {author} {\bibfnamefont {J.~V.}\ \bibnamefont {Porto}},
  \bibinfo {author} {\bibfnamefont {B.~L.}\ \bibnamefont {Tolra}}, \bibinfo
  {author} {\bibfnamefont {J.~M.}\ \bibnamefont {Obrecht}}, \bibinfo {author}
  {\bibfnamefont {B.~E.}\ \bibnamefont {King}}, \bibinfo {author}
  {\bibfnamefont {M.}~\bibnamefont {Subbotin}}, \bibinfo {author}
  {\bibfnamefont {S.~L.}\ \bibnamefont {Rolston}}, \ and\ \bibinfo {author}
  {\bibfnamefont {W.~D.}\ \bibnamefont {Phillips}},\ }\href {\doibase
  10.1103/PhysRevA.67.051603} {\bibfield  {journal} {\bibinfo  {journal} {Phys.
  Rev. A}\ }\textbf {\bibinfo {volume} {67}},\ \bibinfo {pages} {051603}
  (\bibinfo {year} {2003})}\BibitemShut {NoStop}%
\bibitem [{\citenamefont {Bloch}\ \emph {et~al.}(2008)\citenamefont {Bloch},
  \citenamefont {Dalibard},\ and\ \citenamefont
  {Zwerger}}]{blochRevModPhys2008}%
  \BibitemOpen
  \bibfield  {author} {\bibinfo {author} {\bibfnamefont {I.}~\bibnamefont
  {Bloch}}, \bibinfo {author} {\bibfnamefont {J.}~\bibnamefont {Dalibard}}, \
  and\ \bibinfo {author} {\bibfnamefont {W.}~\bibnamefont {Zwerger}},\ }\href
  {\doibase 10.1103/RevModPhys.80.885} {\bibfield  {journal} {\bibinfo
  {journal} {Rev. Mod. Phys.}\ }\textbf {\bibinfo {volume} {80}},\ \bibinfo
  {pages} {885} (\bibinfo {year} {2008})}\BibitemShut {NoStop}%
\bibitem [{\citenamefont {Giorgini}\ \emph {et~al.}(2008)\citenamefont
  {Giorgini}, \citenamefont {Pitaevskii},\ and\ \citenamefont
  {Stringari}}]{giorginiRevModPhys2008}%
  \BibitemOpen
  \bibfield  {author} {\bibinfo {author} {\bibfnamefont {S.}~\bibnamefont
  {Giorgini}}, \bibinfo {author} {\bibfnamefont {L.~P.}\ \bibnamefont
  {Pitaevskii}}, \ and\ \bibinfo {author} {\bibfnamefont {S.}~\bibnamefont
  {Stringari}},\ }\href {\doibase 10.1103/RevModPhys.80.1215} {\bibfield
  {journal} {\bibinfo  {journal} {Rev. Mod. Phys.}\ }\textbf {\bibinfo {volume}
  {80}},\ \bibinfo {pages} {1215} (\bibinfo {year} {2008})}\BibitemShut
  {NoStop}%
\bibitem [{\citenamefont {Read}\ and\ \citenamefont
  {Green}(2000)}]{NReadPRB2000}%
  \BibitemOpen
  \bibfield  {author} {\bibinfo {author} {\bibfnamefont {N.}~\bibnamefont
  {Read}}\ and\ \bibinfo {author} {\bibfnamefont {D.}~\bibnamefont {Green}},\
  }\href {\doibase 10.1103/PhysRevB.61.10267} {\bibfield  {journal} {\bibinfo
  {journal} {Phys. Rev. B}\ }\textbf {\bibinfo {volume} {61}},\ \bibinfo
  {pages} {10267} (\bibinfo {year} {2000})}\BibitemShut {NoStop}%
\bibitem [{\citenamefont {Liu}\ and\ \citenamefont {Yin}(2012)}]{BLiuPRA2012}%
  \BibitemOpen
  \bibfield  {author} {\bibinfo {author} {\bibfnamefont {B.}~\bibnamefont
  {Liu}}\ and\ \bibinfo {author} {\bibfnamefont {L.}~\bibnamefont {Yin}},\
  }\href {\doibase 10.1103/PhysRevA.86.031603} {\bibfield  {journal} {\bibinfo
  {journal} {Phys. Rev. A}\ }\textbf {\bibinfo {volume} {86}},\ \bibinfo
  {pages} {031603} (\bibinfo {year} {2012})}\BibitemShut {NoStop}%
\bibitem [{\citenamefont {Han}\ \emph {et~al.}(2009)\citenamefont {Han},
  \citenamefont {Chan}, \citenamefont {Yi}, \citenamefont {Daley},
  \citenamefont {Diehl}, \citenamefont {Zoller},\ and\ \citenamefont
  {Duan}}]{YJHanPRL2009}%
  \BibitemOpen
  \bibfield  {author} {\bibinfo {author} {\bibfnamefont {Y.-J.}\ \bibnamefont
  {Han}}, \bibinfo {author} {\bibfnamefont {Y.-H.}\ \bibnamefont {Chan}},
  \bibinfo {author} {\bibfnamefont {W.}~\bibnamefont {Yi}}, \bibinfo {author}
  {\bibfnamefont {A.~J.}\ \bibnamefont {Daley}}, \bibinfo {author}
  {\bibfnamefont {S.}~\bibnamefont {Diehl}}, \bibinfo {author} {\bibfnamefont
  {P.}~\bibnamefont {Zoller}}, \ and\ \bibinfo {author} {\bibfnamefont {L.-M.}\
  \bibnamefont {Duan}},\ }\href {\doibase 10.1103/PhysRevLett.103.070404}
  {\bibfield  {journal} {\bibinfo  {journal} {Phys. Rev. Lett.}\ }\textbf
  {\bibinfo {volume} {103}},\ \bibinfo {pages} {070404} (\bibinfo {year}
  {2009})}\BibitemShut {NoStop}%
\bibitem [{\citenamefont {Ho}\ and\ \citenamefont
  {Diener}(2005)}]{TLHoPRL2005}%
  \BibitemOpen
  \bibfield  {author} {\bibinfo {author} {\bibfnamefont {T.-L.}\ \bibnamefont
  {Ho}}\ and\ \bibinfo {author} {\bibfnamefont {R.~B.}\ \bibnamefont
  {Diener}},\ }\href {\doibase 10.1103/PhysRevLett.94.090402} {\bibfield
  {journal} {\bibinfo  {journal} {Phys. Rev. Lett.}\ }\textbf {\bibinfo
  {volume} {94}},\ \bibinfo {pages} {090402} (\bibinfo {year}
  {2005})}\BibitemShut {NoStop}%
\bibitem [{\citenamefont {Iskin}\ and\ \citenamefont
  {de~Melo}(2005)}]{iskinPhysRevB2005}%
  \BibitemOpen
  \bibfield  {author} {\bibinfo {author} {\bibfnamefont {M.}~\bibnamefont
  {Iskin}}\ and\ \bibinfo {author} {\bibfnamefont {C.~A. R.~S.}\ \bibnamefont
  {de~Melo}},\ }\href {\doibase 10.1103/PhysRevB.72.224513} {\bibfield
  {journal} {\bibinfo  {journal} {Phys. Rev. B}\ }\textbf {\bibinfo {volume}
  {72}},\ \bibinfo {pages} {224513} (\bibinfo {year} {2005})}\BibitemShut
  {NoStop}%
\bibitem [{\citenamefont {Massignan}\ \emph {et~al.}(2010)\citenamefont
  {Massignan}, \citenamefont {Sanpera},\ and\ \citenamefont
  {Lewenstein}}]{massignanPhysRevA2010}%
  \BibitemOpen
  \bibfield  {author} {\bibinfo {author} {\bibfnamefont {P.}~\bibnamefont
  {Massignan}}, \bibinfo {author} {\bibfnamefont {A.}~\bibnamefont {Sanpera}},
  \ and\ \bibinfo {author} {\bibfnamefont {M.}~\bibnamefont {Lewenstein}},\
  }\href {\doibase 10.1103/PhysRevA.81.031607} {\bibfield  {journal} {\bibinfo
  {journal} {Phys. Rev. A}\ }\textbf {\bibinfo {volume} {81}},\ \bibinfo
  {pages} {031607} (\bibinfo {year} {2010})}\BibitemShut {NoStop}%
\bibitem [{\citenamefont {Shirley}(1965)}]{shirleyPhysRev1965}%
  \BibitemOpen
  \bibfield  {author} {\bibinfo {author} {\bibfnamefont {J.~H.}\ \bibnamefont
  {Shirley}},\ }\href {\doibase 10.1103/PhysRev.138.B979} {\bibfield  {journal}
  {\bibinfo  {journal} {Phys. Rev.}\ }\textbf {\bibinfo {volume} {138}},\
  \bibinfo {pages} {B979} (\bibinfo {year} {1965})}\BibitemShut {NoStop}%
\end{thebibliography}%

\end{document}